\documentstyle[aps,preprint]{revtex}
\tightenlines
\begin{document}
\draft
\title {Vanishing spin alignment : experimental indication of triaxial
$\bf ^{28}Si + {^{28}Si}$ nuclear molecule }
 
\author { R. Nouicer$^{1,}$\thanks{Present address: University of Illinois at
Chicago and Argonne National Laboratory, IL 60439, USA.}, C. Beck$^{1}$, R.M. 
Freeman$^{1}$, F. Haas$^{1}$, N. Aissaoui$^{1}$, T. Bellot$^{1}$, G. de
France$^{1,}$\thanks{Present address: GANIL, Caen, France.}, D. Disdier$^{1}$,
G. Duch\^ene$^{1}$, A. Elanique$^{1,3}$, A. Hachem$^{1}$, F. Hoellinger$^{1}$,
D. Mahboub$^{1,3}$, V. Rauch$^{1}$, S.J. Sanders$^{4}$, A. Dummer$^{4}$, F.W.
Prosser$^{4}$, A. Szanto de Toledo$^{2}$, Sl. Cavallaro$^{3}$, E. Uegaki$^{5}$
and Y. Abe$^{6}$} 

\address{\it $^{1}$ IReS, UMR7500, CNRS-IN2P3 et Universit\'e Louis Pasteur,
B.P.28, F-67037 Strasbourg, Cedex 2, France } 
\address {\it $^{2}$ Instituto de Fisica da Universidade de S\~ao Paulo,
S\~ao Paulo, Brazil} 
\address {\it $^{3}$ Dipartimento di Fisica dell'Universit\'a di Catania, INFN
and LNS Catania, Italy } 
\address {\it $^{4}$ Department of Physics and Astronomy, University of Kansas,
Lawrence, KS 66045, USA }
\address {\it $^{5}$ Department of Physics, Akita University, Akita 010, Japan}
\address {\it $^{6}$ Yukawa Institute for Theoretical Physics, Kyoto
University, Kyoto, Japan } 

\date{\today}
\maketitle

\newpage

\begin{abstract}
{Fragment-fragment-$\gamma$ coincidences have been measured for $\rm ^{28}Si +
{^{28}Si}$ at an energy corresponding to the population of a conjectured
resonance in $^{56}$Ni. Fragment angular distributions as well as $\gamma$-ray
angular correlations indicate that the spin orientations of the outgoing
fragments are perpendicular to the orbital angular momentum. This differs from
the $\rm ^{24}Mg+{^{24}Mg}$ and the $\rm ^{12}C+{^{12}C}$ resonances, and
suggests two oblate $\rm ^{28}Si$ nuclei interacting in an equator-to-equator
molecular configuration.} 
\end{abstract} 

\vskip 2.0cm
 

\newpage

In heavy-ion collisions the observations of unusual modes of nuclear
excitations, such as giant dipole resonances built on excited states, scissors
mode vibrations, and quasimolecular resonances, have led to important insights
regarding nuclear and subnuclear degrees of freedom. Most of these special
dynamical modes can be understood as collective oscillations around potential
minima in the macroscopic nuclear potential energy surface. These minima, which
may correspond to spherical, deformed, and superdeformed configurations of the
composite system, can allow states which are sufficiently long-lived to
strongly influence the dynamics of the system. The search for nuclear molecules
has a long history \cite{Bet97} starting with the pionnering discovery
\cite{Bro60} of the so-called quasimolecular resonances in the $\rm ^{12}C
+{^{12}C}$ scattering in the Coulomb barrier region. Subsequently, intermediate
width resonances were also discovered in the excitation functions of mutual
inelastic $\rm ^{12}C +{^{12}C}$ scattering yields well above the Coulomb
barrier \cite{Bet97,Kon85}. These latter resonances were found to be associated
with a mutually aligned component \cite{Kon85} suggestive of the formation of
a rotating dinuclear complex in an equator-equator sticking configuration
\cite{Sch96} because of the oblate shape of $\rm ^{12}C$. The most intriguing
evidence for such exotic excitation modes in dinuclear systems is the
observation of pronounced, narrow, and well-isolated resonant structures in
elastic and inelastic excitation functions measured for various medium-mass
compound nuclei (CN) ($40 \le A_{\rm CN} \le 60$) \cite{Bet81,Zur83}. The
observation of resonant structures in the medium-mass region was first reported
for the $\rm ^{28}Si + {^{28}Si}$ reaction \cite{Bet81}, and subsequently for
the $\rm ^{24}Mg + {^{24}Mg}$ reaction \cite{Zur83,Wuo90}. This resonant
structure, strongly correlated in the exit channels, suggested a correspondance
to quasimolecular states in $\rm ^{56}Ni$ at high excitation energy
($E^{*}_{\rm CN}= 60$--$75\rm\,MeV$) and high angular momenta
($34$--$42\,\hbar$). These large values of angular momenta are of special
interest because they exceed the rotating liquid drop model limit \cite{Wuo90}.

\par

Based on the results of Nilsson-Strutinsky calculations, it has been suggested
\cite{Bet85} that shell-stabilized superdeformed states may exist in the
secondary minima of the adiabatic potential energy surfaces for the $\rm
^{56}Ni$ nucleus in the region of high $E^{*}_{\rm CN}$ and large angular
momenta relevant to the observed resonances. The $\rm ^{28}Si + {^{28}Si}$
resonances would then be associated with metastable quasimolecular
configurations with extreme deformations. Spin alignment measurements
\cite{Wuo87,Mat87} for the resonant $\rm ^{24}Mg+{^{24}Mg}$ system are already
available. Based on these measurements, a deformed configuration is suggested
for the $^{48}$Cr dinuclear system that corresponds to two prolate deformed
$^{24}$Mg nuclei in a pole-to-pole arrangement \cite{Wuo90}. Because of the
complexity of the resonant structure, where several narrow resonances are found
to have the same resonance spin, its analysis solely within a static approach
is difficult. Dynamical aspects of this dinuclear complex were studied within a
molecular approach \cite{Maa88,Ueg89}. Similar calculations have been applied
for $\rm ^{28}Si+{^{28}Si}$ \cite{Ueg94} which is, however, an axially 
nonsymmetric system arising from the oblate deformation of the $^{28}$Si nucleus
in its ground state. To explore these differences and to obtain more precisely
the triaxial properties of the $\rm ^{28}Si+{^{28}Si}$ resonances, we have
performed a high-statistics experimental study of the $\rm ^{28}Si+{^{28}Si}$
collision at an energy corresponding to a conjectured $J^{\pi}=38^{+}$
resonance in $^{56}$Ni \cite{Bet81}. 

\par

In this Rapid Communication we report on experimental results obtained at the
Strasbourg {\sc VIVITRON} Tandem accelerator using a $^{28}$Si beam of energy
$E_{\rm lab}=111.6\rm\,MeV$. The analyzing magnet was calibrated before and
after the experiment with a reproducibility of better than $0.07 \%$ to make
sure that the chosen energy does well populate the resonance \cite{Bet81}. The
beam struck a $25\rm\,{\mu}g/cm^{2}$ thick $^{\rm nat}$Si ($92 \%$ of $\rm
^{28}Si$) target. The Si target thickness corresponds to a beam energy loss of
$\Delta E = 130\rm\,keV$, which is smaller than the width of the resonance
(${\Gamma}_{\rm lab} \approx 300\rm\,keV$). In order to check the beam energy,
previous results of large-angle elastic scattering cross sections \cite{Dic80}
available at close bombarding energies were compared with the present
experiment. Heavy fragments were detected in coincidence using two large-area
Si position-sensitive detectors located symmetrically on either side of the
beam axis in order to cover the laboratory angular range from 22$^{\circ}$ to
73$^{\circ}$ in the horizontal plane with a solid angle of 114 msr and a
vertical angular acceptance of $\approx$ $\pm$ 4$^{\circ}$. The $\gamma$ rays
emitted from the fragments were detected in the 54 Compton-suppressed Germanium
detectors of the {\sc EUROGAM Phase II} multidetector array. Data were taken
in both double and triple coincidence modes : fragment-fragment (F-F) and
fragment-fragment-$\gamma$ (F-F-$\gamma$), respectively. 

\par

The excitation-energy ($E_{\sc x}$) spectrum corrected for F-F detection
efficiency in the $\rm ^{28}Si + {^{28}Si}$ exit-channel is shown as
histograms in Fig.~1. The cross sections have been obtained after integration
over the range of scattering angles $61^{\circ} \le \theta^{\rm FF}_{\rm c.m.}
\le 114^{\circ}$. The spectrum is dominated by peaks corresponding to
transitions to large negative $Q$-values and to mutual inelastic excitations
with large spins. For $E_{\sc x} \leq 10\rm\,MeV$ the identification of most of
the states based on their $\gamma$ decay is straightforward and spin
assignments of the lowest energy peaks are given in the figure. Their peak
yields are in very good agreement with previous elastic, 2$^{+}$, and
(2$^{+}$,2$^{+}$) cross sections data \cite{Bet81,Bet97b} taken at the 38$^{+}$
resonance energy of $E_{\rm lab}=112\rm\,MeV$. An analysis of the results of
Ref.\cite{Dic80} taken at three bombarding energies 111, 112, and $113\rm\,MeV$,
which span the resonance in the same angular range $90^{\circ} \pm
1.8^{\circ}$, gives the ratios of the mutual to single 2$^{+}$ excitations as
$\displaystyle{\sigma(2^{+},2^{+})/ \sigma (2^{+})}=1.45 \pm 0.01$, $1.64 \pm
0.02$ and $1.24 \pm 0.01$, respectively. The present data yield
$\displaystyle{\sigma(2^{+},2^{+})/ \sigma (2^{+})}= 1.65 \pm 0.01$ indicates
that the VIVITRON beam energy chosen for this work was close to the resonance
energy of 112 MeV. The dashed line in Fig.~1 shows the results of
statistical-model calculations according to the transition-state model ({\sc
TSM}) for fission decay using standard parameters for $^{56}$Ni CN
\cite{San99}. At high excitation energies (E$_{\sc x} \ge 10 \rm\,MeV)$, where
the number of possible mutual excitations is large, the {\sc TSM} calculations
are in qualitative agreement with the data and suggest the importance of the
fission decay in this energy region. For discrete low-lying states ($E_{\sc x}
\le 4 \rm\,MeV$) a resonant, nonstatistical behavior is known to dominate the
yields of these exit-channels and, consequently, as shown in Fig.~1, it is not
surprising that the {\sc TSM} calculations fail to describe these yields. 

\par

In order to investigate the resonant effects and to determine their most
favorable angular momentum, we extracted the F-F angular distributions (AD) by
selecting excitation energy ranges corresponding to different states of the two
$\rm ^{28}Si$ fragments \cite{Nou98,Bec99}. For the low-lying states, the
various (elastic, $2^{+}$, $2^{+}$-$2^{+}$) exit-channels were found to display
strongly oscillatory AD at backward angles in the center-of-mass (c.m.) range
$68^{\circ}\le \theta^{\rm FF} _{\rm c.m.}\le 111^{\circ}$ as shown in Fig.~2.
Although very weak at backward angles the elastic cross sections are some 2 to
3 orders of magnitude above optical model predictions \cite{Bet85}. The
$2^{+}_{1}$-$2^{+}_{1}$ group could not be resolved from the much weaker
excited $4^{+}_{1}$ level in the F-F data alone. However, the F-F-$\gamma$ data
were used to determine the position of the $4^{+}_{1}$ state and to set a
narrow $2^{+}_{1}$-$2^{+}_{1}$ coincidence gate in Fig.~1 accordingly to
minimize the $4^{+}_{1}$ contribution \cite{Nou98,Bec99}. The $\gamma$-ray
efficiency corrected F-F-$\gamma$ spectrum gated by the $4^{+}_{1}$ to
$2^{+}_{1}$ $\gamma$-ray transition (E$_{\gamma}$ = 2839 keV) \cite{Nou98}
(shaded histograms) shows that only a very small fraction of the
$2^{+}_{1}$-$2^{+}_{1}$ peak (it can be estimated to be less than 5 $\%$)
correspond to a contribution of the $4^{+}_{1}$ state in the F-F spectrum of
Fig.1. Similarly narrow Q-value gates were chosen for the elastic and single
$2^{+}$ peaks (a Gaussian fit of the $2^{+}$ peak would suggest a contribution
of less than 6 $\%$ to the mutual peak). The regular oscillations are periodic
and thus rather well described by the solid curves of Fig.~2 calculated with
[P$_{L=38}$(cos$\theta$)]$^{2}$ (the dashed and dotted-dashed lines have been
calculated with L=36 and L=40 respectively). Despite the possible contamination
of more direct reaction mechanisms and the admixture of smaller angular
momenta, $L=38$ gives clearly a superior fit (at least for the elastic and
single inelastic channels) in good agreement previous elastic measurements
\cite{Bet81,Dic80}. The $2^{+}_{1}$-$2^{+}_{1}$ channel might also contain the
contribution of $L=36$ but the $L=40$ cannot be completely ruled out either.
The AD's measured for the four lowest energy states are consistent with AD
results obtained at the resonance energy for the same channels \cite{Bet97b}.
The fact that the measured AD's correspond to shapes characterized by the same
angular dependence indicates that the resonant behavior is favored by the
partial wave associated with $L= 38\,\hbar$. Although this is not a firm spin
assignment, it does suggest that this might be the spin value of the resonance
in accordance with the previous claims \cite{Bet81,Bet97b}. Since the total
angular momentum $\vec{J} = \vec{L} + \vec{S}$ (where $\vec{S}$ represents the
total channel spin of the fragments) is conserved and $L =38\,\hbar$ is the
most favorable partial wave in the three resonant exit-channels, this implies
that the projection of the fragment spins along the direction perpendicular to
the reaction plane corresponding to the magnetic substate $m=0$. For higher
excitation energy $E_{\sc x} \ge 6\rm\,MeV$, the identification of the dominant
excited states (4$_{1}^{+}$, 2$_{1}^{+}$) and (4$_{1}^{+}$,~4$_{1}^{+}$) shown
in Fig.~1 was verified by using the F-F-$\gamma$ data \cite{Nou98,Bec99}. In
Fig.~2, the AD for the mutual excitation states (4$_{1}^{+}$, 2$_{1}^{+}$) at
an excitation energy $E_{\sc x}\approx 6.7\rm\,MeV$ is not as strongly
structured as the AD for the low-lying states, whereas the AD for the mutual
excited states (4$_{1}^{+}$,~4$_{1}^{+}$) has a shape comparable to
1/$\sin\theta^{\rm FF}_{\rm c.m.}$ indicating that the major part of the yields
might have a dominant statistical fission origin as predicted by the TSM
predictions (see Fig.~1). 

\par 

The following analysis is focussed on the F-F-$\gamma$ data in the $\rm ^{28}Si
+ {^{28}Si}$ exit channel with both $\rm ^{28}Si$ fragments being detected in
$\theta^{\rm FF}_{\rm c.m.}=90^{\circ} \pm 7^{\circ}$ (see Fig.~2) with a very
narrow coincident gate set on the 2$^{+}_{1}$-2$^{+}_{1}$ peak of Fig.~1. Three
quantization axes have been defined as follows~:~(a) the beam axis, (b) the
axis normal to the scattering plane, and (c) the axis perpendicular to (a) and
(b) axes. Since the two $^{28}$Si fragments are detected in the angular region
83$^{\circ}$~$\le \theta^{\rm FF}_{\rm c.m.}$~$\le$~97$^{\circ}$, the (c) axis
corresponds approximately to the molecular axis of the out-going binary
fragments. The experimental results of the $\gamma$-ray angular correlation
$W(\theta_\gamma)$ (the polar $\theta_{\gamma}$ and azimuthal $\phi_{\gamma}$
angles being defined relatively to each of the quantization axes) for the
2$^{+}_{1}$-2$^{+}_{1}$ exit-channel are shown in Fig.~3 by points. The mutual
excitation channel is here presented rather than the single excitation channel
since it is known to induce more spin alignment in resonant systems such as
$\rm ^{12}C + {^{12}C}$ \cite{Kon85} and $\rm ^{24}Mg + {^{24}Mg}$
\cite{Mat87}. The analysis method of the $W(\theta_\gamma)$ data is described
in Ref.\cite{Wuo90}, in which the process of integration over $\phi_{\gamma}$
requires, due to the geometry of the {\sc EUROGAM} spectrometer, some averaging
over $\theta_{\gamma}$. Our experimental efficiency for F-F-$\gamma$-$\gamma$
detection is low and, consequently, the analysis was done with the condition of
a $\gamma$ multiplicity equal to one. The strong minimum in (b) at 90$^{\circ}$
implies $m=0$ (see the following discussion) and thus that the intrinsic spin
vectors of the 2$^{+}$ states are oriented in the reaction plane
perpendicularly to the orbital angular momentum. The value of the total orbital
angular momentum therefore remains close to $L = 38\,\hbar$, in good agreement
with the AD results. The 4$\pi$ geometry of the $\gamma$-ray spectrometer also
allowed us to fit the $W(\theta_\gamma)$ (see Fig.~3, solid curve) to obtain
more quantitative information about the contributions from the different
magnetic substates (see Table~\ref{axe}). As proposed in \cite{Wuo90}, the
$W(\theta_\gamma)$ data have been described by an expression of the form
$W(\theta_\gamma) = \sum_{m}^{} P_{m}W_{m}(\theta_\gamma)$, where $P_{m}$
represent either the magnetic substate population parameters in single or
mutual 2$^{+}$ inelastic scattering, or the relative intensities of transitions
with different $\Delta m$ for E2 transitions between higher excited states.
Since the parameters $P_{m}$ enter the expression as linear coefficients of the
pure-$\Delta m$ functions $W_{\Delta m}(\theta_\gamma)$, the fits were
calculated using a simple linear least squares procedure \cite{Wuo90}. The fit
values given in Table~\ref{axe} show a significant {\it m=0} substate
population, favored for both quantization axes (a) and (b) but not in (c),
consistent with the spin vectors oriented in the reaction plane perpendicularly
to the total angular momentum. The significant contributions of m = $\pm$ 2 are
consistent with the AD analysis of the $2^{+}_{1}$-$2^{+}_{1}$ which contain
contributions from $L$=36 and 40. Admixtures of these other smaller
orientations do not affect the observation of the dominant ``disalignment"
component consistent with the AD results. 

\par

The present AD and $W(\theta_\gamma)$ results are in contrast with the
alignment found for the $\rm ^{24}Mg+{^{24}Mg}$ reaction \cite{Wuo87,Mat87}.
The reason why the two systems differ so strongly may be associated with the
structure of the dinucleus configuration. In $\rm ^{24}Mg+{^{24}Mg}$, the
energetically favored configuration is the pole-to-pole configuration (see
Fig.~1 of Ref.\cite{Ueg89}) due to the prolate shape of $^{24}$Mg. The system
rotates about a minor axis of the $^{24}$Mg nuclei (perpendicular to the
symmetry axis). Therefore, the total angular momentum is parallel to the
$^{24}$Mg spins to give rise to alignment. The energetically favored
configuration of an oblate-oblate dinuclear system is the equator-equator
configuration (see Fig.~1 of Ref.\cite{Ueg94}), with two pancakes touching each
other side-to-side ($\rm ^{28}Si$ is oblate in its ground state and the feeding
of the bands of $^{28}$Si has revealed that this nucleus is dominated by states
with {\it oblate} deformation \cite{Nou98,Bec99}). At a given angular momentum
$J$, this configuration rotates in a triaxial way approximately about the axis
normal to the plane defined by the two pancake-like nuclei which corresponds to
the largest moment of inertia in the state with the lowest energy \cite{Ueg99}.
The spins of the $^{28}$Si fragments are thus in this plane ($m$=0) since no
rotation can occur about their symmetry axes. This analysis is consistent with
the conclusions obtained from the analysis of the AD data. The molecular-model
calculations \cite{Ueg94} have been developed \cite{Ueg99} to take into account
the fact that the largest moment of inertia $I_{\it x}$ is only slightly larger
than the moment of inertia $I_{\it y}$ about the in-plane axis that is
perpendicular to the relative vector. This means that the total system is
slightly axial asymmetric about the {\it z}-axis and therefore, the angular
momentum vector is not completely parallel to the {\it x}-axis. To obtain an
accurate description of this triaxial rotor, as it is wellknown for polyatomic
molecules, we diagonalize the Hamiltonian of an asymmetric inertia tensor,
which gives rise to a mixing of the $K$ projections of the total spin $J$. In
the high-spin limit ($K/J$ $\approx$ 0), the diagonalization is found to be
equivalent to solving a differential equation of the harmonic oscillator with
parameters given by the moment of inertia \cite{Ueg99,Boh75}. With the wave
function obtained for the lowest energy state we have calculated the P$_{m}$'s
(given in Table I) as well as the $W(\theta_\gamma)$ (dashed lines) for the
mutual inelastic channel (for which the second unobserved $\gamma$-ray is
summed over all possible directions) which are compared with the data in
Fig.~3. The characteristic features of the experiment (points) are in agreement
with the molecular-model predictions \cite{Ueg99}. A similar theoretical 
investigation was independently proposed to describe the $\rm ^{12}C +
{^{12}C}$ scattering as an oblate-oblate dinuclear system in equator-equator
orientations \cite{Sch96}. Although other possible pictures might be equally
consistent to explain both the $\rm ^{28}Si + {^{28}Si}$ and $\rm ^{24}Mg +
{^{24}Mg}$ scatterings, the present data will put severe constraints on future
attempts with alternate model descriptions, such as, the double resonance model
\cite{Lan82} or other coupled channel calculations \cite{Thi84}. 

\par

In summary, the present high-resolution study of fragment-fragment-$\gamma$
coincidence data collected with a powerful 4$\pi$ $\gamma$-ray spectrometer for
$\rm ^{28}Si + {^{28}Si}$ at $E_{\rm lab}=111.6\rm\,MeV$, populating a well
known molecular resonance in $\rm ^{56}Ni$, does not show as strong evidence of
fragment spin alignment with respect to the orbital angular momentum as found
previously for $\rm ^{12}C + {^{12}C}$ and $\rm ^{24}Mg + {^{24}Mg}$. This was
first deduced from the measured fragment-fragment angular distributions of the
elastic~0$^{+}_{1}$, inelastic 2$^{+}_{1}$, and mutual inelastic
2$^{+}_{1}$-2$^{+}_{1}$ channels, which appear to be rather well described by a
partial wave with $L=38\,\hbar$, and was confirmed by the analysis of the
fragment-fragment-$\gamma$ angular correlations for the mutual inelastic
channel. These observations different from that of a spin alignment evidenced for
$\rm ^{24}Mg+ {^{24}Mg}$ resonances may support the occurence, predicted by the
molecular model, of a stable $\rm ^{28}Si+ {^{28}Si}$ oblate-oblate dinuclear
system in which the equator-equator spin orientations result from its triaxial
configuration. Similar high-statistics exclusive measurements at energies which
do not correspond to a molecular resonance would be very instructive. 

\vskip 1.4cm

{\small
\noindent
We are pleased to thank the {\sc VIVITRON} operators and the {\sc EUROGAM}
staff of IReS. R. R. Betts and A. H. Wuosmaa are warmly acknowledged for a
careful reading of the manuscript. This work was sponsored by the C.N.R.S.
within CNRS/NSF and CNRS/CNPq Collaboration programs and also in part by the
U.S. DOE under Grant No. DE-FG03-96ER40481. }

\begin{table}[hpt]
\caption{\label{axe}
Magnetic substate population parameters deduced from fits corresponding to each
quantization axis of Fig.~3 and from the model predictions discussed in the
text.} 
\begin{tabular}[t]{lcccccc}
Quantization
&\multicolumn{2}{c}{$P_{m=0}$}&
\multicolumn{2}{c}{P$_{m=\pm 1}$}&
\multicolumn{2}{c}{P$_{m=\pm 2}$}\\
\cline{2-7}
 \qquad axis&Fit&Model&Fit&Model&Fit&Model\\
\hline
\qquad(a)  &0.30 $\pm$ 0.08 &0.48&0.16 $\pm$0.04 &0.17&0.18$\pm$0.05 &0.09\\
\qquad(b)  &0.46 $\pm$ 0.05 &0.48&0              & 0 &0.27$\pm$0.02 &0.26\\
\qquad(c)  &0.14 $\pm$ 0.05 &0.02&0.17 $\pm$0.03 &0.17&0.26$\pm$0.04 &0.32\\
\end{tabular}
\end{table}

\begin{figure}

\caption{Excitation-energy spectra for the $\rm ^{28}Si + {^{28}Si}$
exit-channel in the angular region $61^{\circ} \le \theta^{\rm FF}_{\rm c.m.}
\le 114^{\circ}$. The efficiency corrected F-F data (histograms) and the {\sc
TSM} calculations (dashed line) are integrated cross sections. The $\gamma$-ray
efficiency corrected F-F-$\gamma$ spectrum gated by the 2839 keV (4$^{+}$
$\rightarrow$ 2$^{+}$) $\gamma$-ray transition is presented as shaded
histograms.} 
\label{fig1}
\end{figure}

\begin{figure}
\caption{Experimental F-F AD of the elastic ($E_{\sc x} \approx 0\rm\,MeV$),
inelastic ($E_{\sc x} \approx 1.7\rm\,MeV$), mutual inelastic ($E_{\sc x}
\approx 3.6\rm\,MeV$) and higher excitations ($E_{\sc x} \approx 6.7$ and
$9.7\rm\,MeV$). The dashed, solid, and dotted-dashed curves represent squared
Legendre polynomials with $L$ = 36, 38, and 40 respectively. The dotted curve
corresponds to a 1/$\sin \theta^{\rm FF}_{\rm c.m.}$ behavior.} 
\label{fig2}
\end{figure}

\begin{figure}
\caption{Experimental F-F-$\gamma$ angular correlations of the mutual inelastic
channel ($2^{+}_{1}$,~$2^{+}_{1}$) in the angular region $83^{\circ} \le
\theta^{\rm FF}_{\rm c.m.} \le 97^{\circ}$ for the three quantization axes
defined in the text. The solid and dashed curves are fits of the data and model
predictions, respectively.} 
\label{fig3} 
\end{figure}

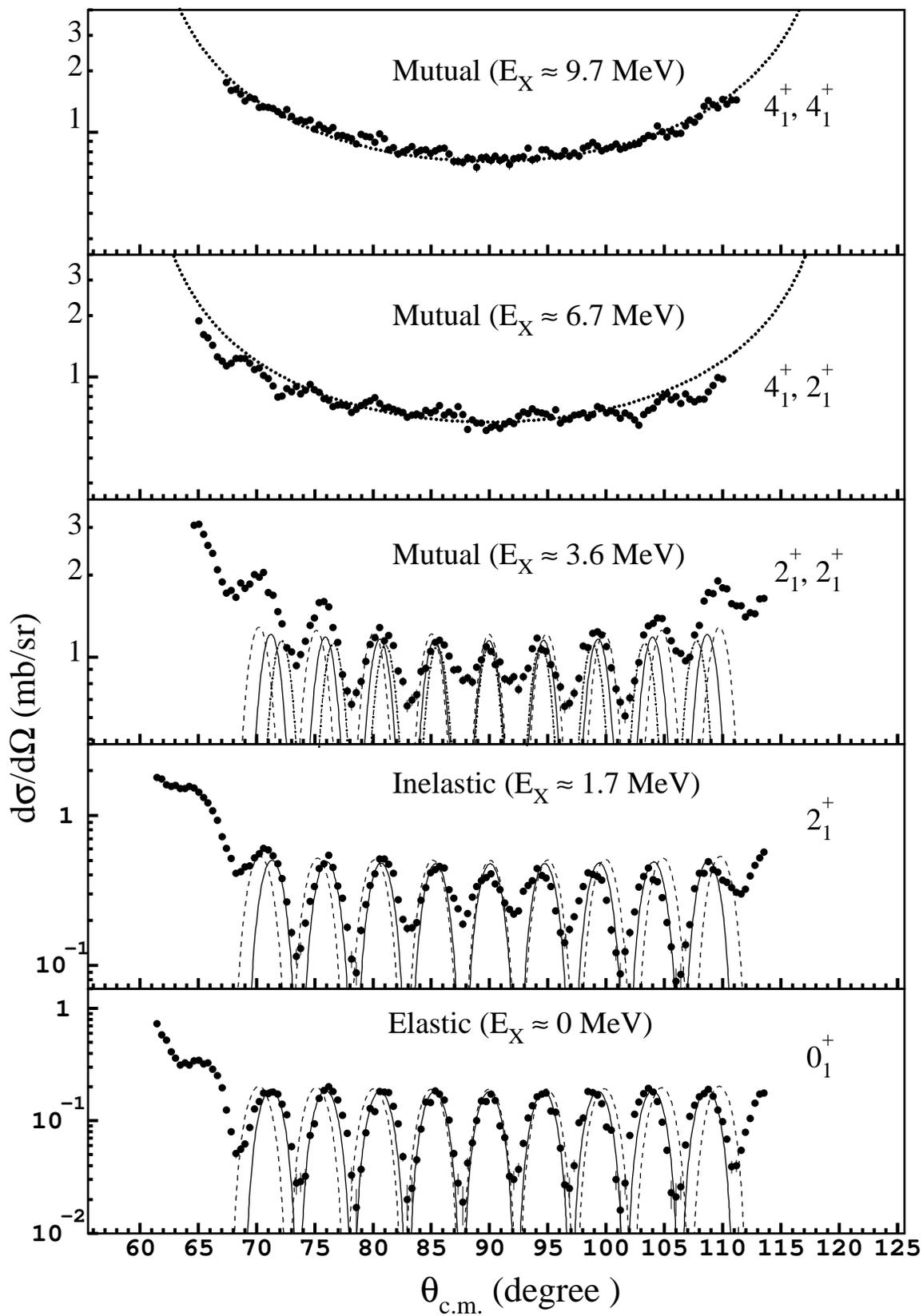
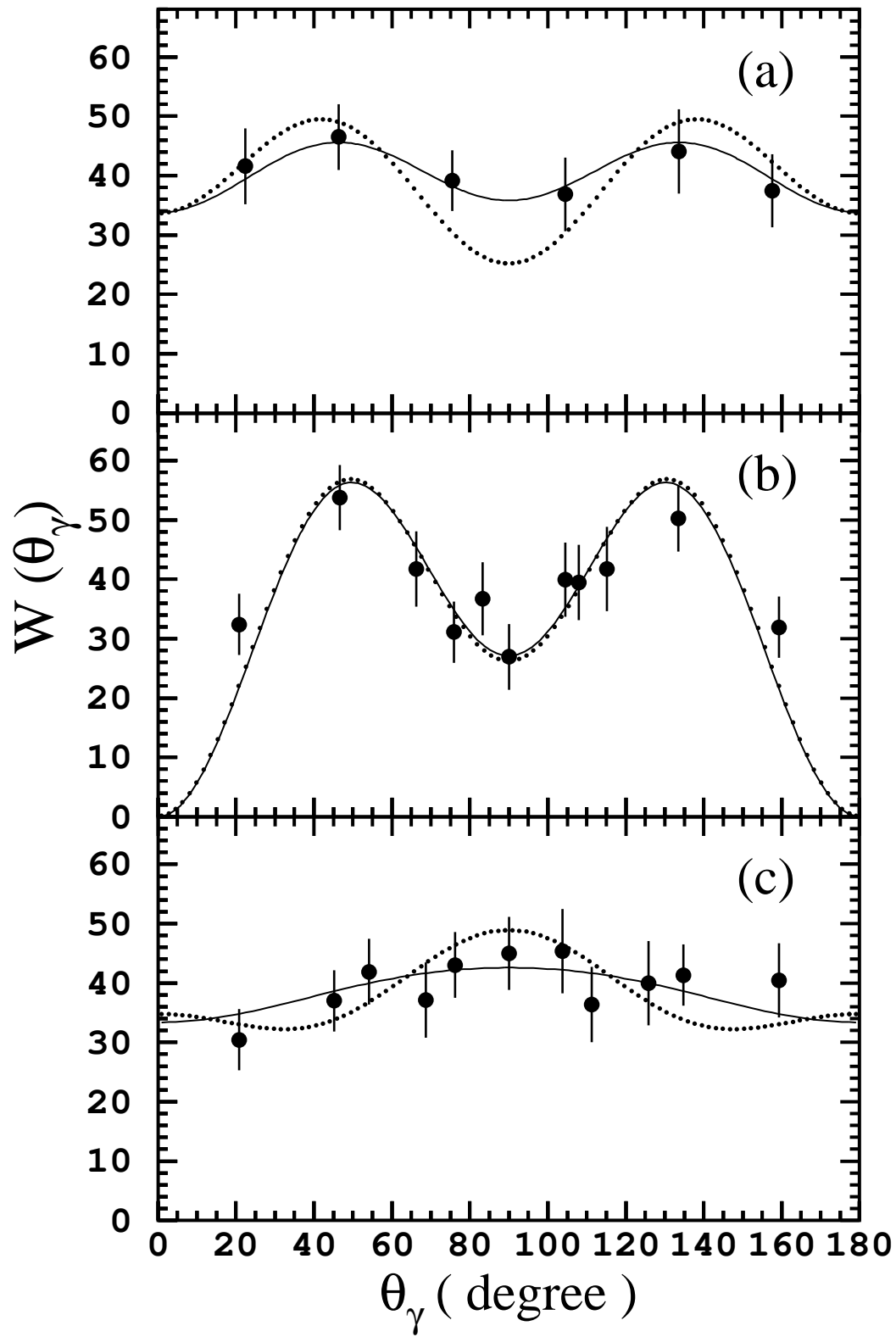
\end{document}